%% file: main.tex
\documentclass[10pt]{llncs}

\usepackage[dvips]{graphicx}
\usepackage{times}
\usepackage{color}
\usepackage{listings}
\usepackage{amsmath}
\usepackage{amssymb}
\usepackage{stmaryrd}
\usepackage[hyphens,spaces]{url}
\usepackage[bookmarks=false]{hyperref}
\usepackage{algorithmicx}
\usepackage{algorithm}
\usepackage{xparse}
\usepackage{wrapfig}
\usepackage{etoolbox}
\usepackage{cite}
\usepackage{inconsolata}
\usepackage{caption}
\usepackage{subcaption}
\usepackage{appendix}

\newtoggle{tech_report}

\toggletrue{tech_report}

\input{macro}

\setMyChor

\begin{document}

\title{\AIOCJ{}: A Choreographic Framework for \\
 Safe Adaptive Distributed Applications}
 \iftoggle{tech_report}{\subtitle{Technical Report}}{}
\author{Mila Dalla Preda\inst{1} \and Saverio 
Giallorenzo\inst{2} \and \\ Ivan Lanese\inst{2} \and Jacopo Mauro\inst{2} 
\and Maurizio Gabbrielli\inst{2} }
\institute{Department of Computer Science - Univ. of Verona
\and Department of Computer Science and Engineering -
 Univ. of Bologna / INRIA
}

\maketitle


\begin{abstract} We present \AIOCJ{}, a framework for programming distributed
adaptive applications. Applications are programmed using AIOC, a choreographic
language suited for expressing patterns of interaction from a global point of
view. AIOC allows the programmer to specify which parts of the application can
be adapted.  Adaptation takes place at runtime by means of rules, which can
change during the execution to tackle possibly unforeseen adaptation needs.
\AIOCJ{} relies on a solid theory that ensures applications to be 
deadlock-free by construction also after adaptation. We describe the 
architecture of \AIOCJ{}, the design of the AIOC language, and an empirical 
validation of the framework. \end{abstract}

\input{intro}
\input{aioc}

\input{implementation}

\input{validation}
\input{related}

\bibliographystyle{ieeetr}
\bibliography{biblio}

\iftoggle{tech_report}{
\clearpage
\appendix
\input{demo}
\input{test_code}
}{}
\end{document}

%% file: macro.tex
\def\AIOCJ{{\textnormal{\textsf{AIOCJ}}}}

\newcommand{\coutLabel}[3]{\overline{#1}\langle#2\rangle@{#3}}

\NewDocumentCommand\scope{mmmmg}{\textsl{scope} \ @#2 \ \{ #3 \} \ 
\textsl{prop} \ \{ #4 \}}
\NewDocumentCommand\pscope{mmmmmg}{#1: \textsl{scope} \ @#3 \ \{ #4 \} \ \textsl{prop} \ \{ #5 
\}
\IfNoValueTF{#6}{}{\ \textsl{roles} \ \{ #6 \}}}

\def\prod{\Pi}

\definecolor{color:keyword}{rgb}{0.53,0.05,0.05}
\definecolor{color:comment}{rgb}{0.25,0.37,0.75}
\definecolor{color:string}{rgb}{0.87,0.0,0.0}
\usepackage{bold-extra}

\lstdefinelanguage{MyChor}{
morekeywords={
  while, if, else, scope, true, false, not, getInput,%
  and, or, prop, from, to, via, on, do, roles, null, include,%
  from, rule, aioc, show, starter, preamble, location
},
sensitive=true,
morecomment=[l]{//},
morecomment=[s]{/*}{*/},
morestring=[b]",
otherkeywords={;,|,:,@,!,N.,E.}
}

\newcommand{\setMyChor}{
\lstset{
  language=MyChor,
  basicstyle=\ttfamily\small
  }
}

\newcommand{\toolplugin}{\textsf{AIOCJ-ecl}}
\newcommand{\toolmid}{\textsf{AIOCJ-mid}}

\lstdefinelanguage{Jolie}{
morekeywords={csets,type,raw,any,undefined,void,default,if,for,while,spawn,
foreach,else,define,main,include,constants,inputPort,outputPort,interface,
execution,cset,nullProcess,RequestResponse,OneWay,throw,throws,
install,scope,embedded,init,synchronized,global,is_defined,is_int,is_bool,
is_long,is_string,bool,long,int,string,double,undef,with,
Location,Protocol,Interfaces,Aggregates,Redirects,linkIn,linkOut},
sensitive=true,
morecomment=[l]{//},
morecomment=[s]{/*}{*/},
morestring=[b]",
otherkeywords={;,|,@}
}

\newcommand{\setJolie}{
	\lstset{
		language=Jolie,
		basicstyle=\ttfamily\small
	}
}

%% file: intro.tex
\section{Introduction}\label{sec:intro} Adaptation is a main feature of
current distributed applications, that should live for a long time in a
continuously changing environment. Anticipating all the possible adaptation
needs when designing an application is very difficult, thus the approaches
able to cope with \emph{unforeseen} adaptation needs are the most interesting. Also, for
distributed applications like the ones that we consider, it is important to
ensure deadlock-freedom (according to~\cite{CONCbugs} about one third   of
concurrency bugs in real applications are deadlocks). While many techniques 
ensuring deadlock freedom exist in the 
literature, e.g., \cite{POPLmontesi,component_deadlock_free,static_deadlock_detection}, 
to the best of our knowledge,  none of them deals with adaptive applications.
Indeed, most of the approaches to adaptation offer no guarantee on
the behaviour of the application after adaptation~\cite{JORBApaper,ADAPT-Bucch
,adapt-distributed01}, or they assume to know all the possible adaptations in
advance~\cite{ZhangGC09}, thus failing to cope with unforeseen adaptation
needs.
%

Here we present \AIOCJ{}, a prototype implementation of a
framework for programming adaptive distributed applications that guarantees
deadlock-freedom by construction 
(the theoretical foundations ensuring this
property are discussed in~\cite{TR}). 
\AIOCJ{} is composed of two parts:
(\emph{i}) a domain-specific language, called Adaptive Interaction-Oriented
Choreographies (\emph{AIOC}) and (\emph{ii}) an \emph{adaptation middleware}
that supports adaptation  of AIOC programs.

The AIOC language describes applications from a global point of view following
the \emph{choreography} paradigm. This paradigm has been applied in different
contexts, see, e.g.,~\cite{hondaESOPext,POPLmontesi,scribble,SEFM08,WSCDL}, but we
are not aware of other tools based on it and targeting adaptive applications.
A choreography defines the interactions among the processes of a distributed
application.  AIOC main innovation consists in two constructs supporting
adaptation: \emph{scopes} and \emph{adaptation   rules}. A scope delimits code
that may be adapted in the future. An adaptation rule provides new code to
replace the one in a given scope. Interestingly, in \AIOCJ{}, adaptation rules
can be defined and inserted in the framework while the application is running,
to cope with adaptation needs which were not foreseen when the application was
designed or even started.


The code below shows a toy AIOC program (left) and an adaptation rule
applicable to it (right).  On the left, Lines 2-4 define a scope in
which the local variable \lstinline{msg} of process \lstinline{user} is
set to \lstinline{"Hello World"}. The keyword \lstinline{prop} defines
properties of the scope (prefixed by \lstinline{N}). In this case, the
\lstinline{name} property is set to \lstinline{"hello_world"}. At Line
5 \lstinline{user} sends the content of \lstinline{msg} to a second
process (\lstinline{display}), that stores it in its local variable
\lstinline{msg}. On the right, Lines 2-3 define the \emph{applicability
  condition} of the rule, i.e., the \lstinline{name} property of the
scope should be set to \lstinline{"hello_world"}, and the
environmental property \lstinline{E.lang} should be equal to
\lstinline{"it"}.  Line 4 shows the code that will replace the one of the
scope, i.e., variable \lstinline{msg} of \lstinline{user} will be set to
\lstinline{"Ciao Mondo"} (Italian for ``Hello World'').

\vspace{-1em}
\begin{minipage}[t]{0.5\columnwidth}
\begin{lstlisting}[mathescape=true,numbers=left,label=es:hello_world,
  basicstyle=\ttfamily\scriptsize]
aioc { 
 scope @user{
  msg@user = "Hello World"
 } prop { N.name = "hello_world"};
 send: user( msg ) -> display( msg ) }
\end{lstlisting}
\end{minipage}
\hspace{1em}
\begin{minipage}[t]{0.5\columnwidth}
\begin{lstlisting}[mathescape=true,numbers=left,label=es:hello_world_rule,
basicstyle=\ttfamily\scriptsize]
rule {
 on { N.name == "hello_world" 
 	and E.lang == "it" } 
 do { msg@user = "Ciao Mondo" }
}
\end{lstlisting}
\end{minipage}

\iftoggle{tech_report}{
An AIOC program describes a full distributed application. \AIOCJ{} generates 
Jolie code for each (possibly distributed)
process.
Jolie~\cite{jolie,joliepaper} is a Service-Oriented orchestration language, 
and we chose it since it provides primitives suitable for implementing our
communication and adaptation behaviour.
}{
	An AIOC program describes a full distributed application. \AIOCJ{} generates, 
	for each distributed process, a service written in Jolie~\cite{jolie,joliepaper}, 
	a Service-Oriented orchestration language.
}

The adaptation middleware consists of a set of adaptation rules stored in
multiple, possibly distributed, \emph{adaptation servers}, and of an
\emph{adaptation manager} that mediates the interactions between the adaptive
application and the various adaptation servers.


\emph{Structure of the paper.} Section~\ref{sec:aioc} presents an overview of
the \AIOCJ{} framework, while Section~\ref{sec:impl} describes its
implementation.  Section~\ref{sec:validation} shows a preliminary validation
of the framework,  with tests on the performances of \AIOCJ{}.
In Section~\ref{sec:conclusions} we discuss related work and future directions
of research. 
\iftoggle{tech_report}{
Appendix \ref{sec:demo} outlines a short demo of the use of the
framework.}{
	A short demo of the use of the framework is available in the companion
	technical report \cite{AIOCJ_TR}.
}

%% file: aioc.tex
\section{Overview: The \AIOCJ{} framework}\label{sec:aioc} 

This section first defines the architectural model that supports adaptation of
\AIOCJ{}  applications and then it introduces the syntax of the AIOC language
via an  example (for a formal presentation of AIOC syntax and  semantics see
\cite{TR}).

\noindent\textbf{The \AIOCJ{} middleware}. We consider applications composed
of processes deployed as services on different localities, including local state and computational 
resources. Each process
has a specific duty in the choreography and follows a given protocol. Processes interact via 
synchronous message passing over channels, also called \emph{operations}. Adaptation is performed by 
an adaptation
middleware including an adaptation manager and some, possibly distributed, adaptation 
servers.
The latter are services that act as repositories of adaptation rules and may
be (manually) added or removed at runtime. Running adaptation
servers register themselves on the adaptation manager. The
running application may interact with the adaptation manager to look for
applicable adaptation rules. The effect of an adaptation rule is to replace a
scope with new code that answers a given adaptation need.
%
%
The adaptation manager checks the rules according to the registration order 
of the adaptation servers, returning the first applicable rule, if any.

\noindent\textbf{The AIOC Language.}
The language relies on a set of roles that identify the processes in
the choreography.
Let us introduce the syntax of the language using an example where Bob invites
Alice to see a film. 

\begin{minipage}{\textwidth}
\begin{lstlisting}[mathescape=true,numbers=left,caption=Appointment 
program.,basicstyle=\ttfamily\scriptsize,label=es:appointment]
include isFreeDay from "calendar.org:80" with http
include getTicket from "cinema.org:8000" with soap
preamble { 
  starter: bob 
  location@bob = "socket://localhost:8000"
  location@alice = "socket://alice.com:8000"
  location@cinema = "socket://cinema.org:8001" }
aioc{ 
  end@bob = false;
  while( ! end )@bob{
    scope @bob {
      free_day@bob = getInput( "Insert your free day" );  
      proposal: bob( free_day ) -> alice( bob_free_day );
      is_free@alice = isFreeDay( bob_free_day );
    } prop { N.scope_name = "matching day" };
    if( is_free )@alice {
      scope @bob {
        proposal: bob( "cinema" ) -> alice( event );
        agreement@alice = getInput( "Bob proposes "  + event +  
          ", do you agree?[y/n]");
        if( agreement == "y" )@alice{
          end@bob = true;
          book: bob( bob_free_day ) -> cinema( book_day );
          ticket@cinema = getTicket( book_day );
          { notify: cinema( ticket ) -> bob( ticket )
            | notify: cinema( ticket ) -> alice( ticket ) }}
      } prop { N.scope_name = "event selection" } };
    if( !end )@bob {
      _r@bob = getInput( "Alice refused. Try another date?[y/n]" );
      if( _r != "y" )@bob{ end@bob = true }}}
\end{lstlisting}
\end{minipage}

The code starts with some deployment information (Lines 1-7), discussed later on. 
The behaviour starts at Line 9.
The program is made by a cycle where 
Bob first checks when
Alice is available and then invites her to the cinema.
Before starting the cycle, Bob initialises the variable 
\lstinline{end}, used in the guard of the cycle, to the boolean
value \lstinline{false} (Line 9). Note 
the annotation \lstinline{@bob}
meaning that \lstinline{end} is local to Bob.
The first
instructions of the while are enclosed in a scope (Lines 11-15),
meaning that they may be adapted in the
future. The first operation within the scope is the call to the
primitive function \lstinline{getInput} that asks to Bob a day where
he is free and stores this date into the local variable
\lstinline{free_day}. At Line 13 the content of \lstinline{free_day} is sent
to Alice via operation \lstinline{proposal}. Alice stores it in
its local variable \lstinline{bob_free_day}.  Then, at Line 14,
Alice calls the external function \lstinline{isFreeDay} that checks
whether she is available on \lstinline{bob_free_day}.
If she is available 
(Line 16) then Bob sends to her the invitation to go to the cinema via
the operation \lstinline{proposal} (Line 18). Alice, reading from the
input, accepts or refuses the invitation (Line 19).
%
If Alice accepts then Bob first sets the variable
\lstinline{end} to \lstinline{true} to end the cycle. Then, he sends to the
cinema the booking request via operation \lstinline{book}. The cinema
generates the tickets using the external function \lstinline{getTicket} and
sends them to Alice and Bob via operation \lstinline{notify}.  The two
notifications are done in parallel using the parallel operator \lstinline{|}
(until now we composed statements using the sequential  operator
\lstinline{;}). Lines 18-26 are enclosed in a second scope with property
\lstinline{N.scope_name = "event selection"}.
If the agreement is not reached, Bob decides, reading 
from the
input, if he wants to stop inviting Alice. If so, the program exits.

We remark the different meanings of the annotations \lstinline{@bob} and 
\lstinline{@alice}. When prefixed by a variable, they identify the owner of 
the variable. Prefixed by the boolean guard of conditionals and cycles, 
they identify the role that evaluates the guard. Prefixed by the keyword
\lstinline{scope}, they identify the process coordinating the
adaptation of that scope.
A scope, besides the code, may also include some properties describing
the current implementation.  These can be specified using the keyword
\lstinline{prop} and are prefixed by \lstinline{N}. For instance,
each scope of the example includes the property
\lstinline{scope_name}, that can be used to distinguish its
functionality.

\AIOCJ{} can interact with external services, seen as functions.  This allows
both to interact with real services and to have easy access to libraries from
other  languages.
To do that, one must specify the address and protocol used to interact
with them. For instance, the external function \lstinline{isFreeDay}
used in Line 14 is associated to the service deployed at the
domain ``calendar.org'', reachable though port $80$, and that uses
http as serialisation protocol (Line 1).
External functions are declared with the keyword \lstinline{include}. To preserve deadlock freedom, 
external services must be non-blocking.
After function declaration, in a \lstinline{preamble} section, it is possible to declare the locations where
processes are deployed. The keyword  \lstinline{starter} is mandatory and
defines which process must be started first. The starter makes
sure all other processes are ready before the execution of the choreography begins.

Now suppose that Bob, during summer, prefers to invite Alice to a
picnic more than to the cinema, provided that the weather forecasts
are good. This can be obtained by adding the following adaptation rule
to one of the adaptation servers. This may even be done while the
application is running, e.g., while Bob is sending an invitation. In
this case, if the first try of Bob is unsuccessful, in the second try he will
propose a picnic.

\begin{minipage}{\textwidth}
\begin{lstlisting}[mathescape=true,numbers=left,caption=Event selection 
adaptation rule., basicstyle=\ttfamily\scriptsize, label=es:appointmentrule]
rule {
  include getWeather from "socket://localhost:8002"
  on { N.scope_name == "event selection" and E.month > 5 and E.month < 10 }
  do { forecasts@bob = getWeather( free_day );
    if( forecasts == "Clear" )@bob{
      eventProposal: bob( "picnic" ) -> alice( event )
    } else { eventProposal: bob( "cinema" ) -> alice( event ) };
    agreement@alice = getInput( "Bob proposes "  + event +  
      ", do you agree?[y/n]");
    if( agreement == "y" )@alice {
      end@bob = true |
      if( event == "cinema" )@alice { 
        //cinema tickets purchase procedure
}}}}
\end{lstlisting}
\end{minipage}

A rule specifies its applicability condition and the new code to execute. In general, the 
applicability condition may depend only on
properties of the scope, environment variables, and variables belonging to the
coordinator of the scope.
In this case, the condition, introduced by the keyword \lstinline{on} (Line 3), makes the
rule applicable to scopes having the property
\lstinline{scope_name} equal to the string  \lstinline{"event selection"} and 
only during summer. This last check relies on an environment variable
\lstinline{month} that contains the current month.
\iftoggle{tech_report}{%
  Remarkably, one can argue that the rule is correct only in the northern 
  hemisphere. However, when the application runs in the southern hemisphere, 
  a different adaptation server will be available, containing a rule
  tailored for the southern hemisphere.
}{}
Environment variables are prefixed by \lstinline{E}.

When the rule applies, the new code to execute is defined using the
keyword \lstinline{do} (Line 4). In this case, the forecasts can be
retrieved calling an external function \lstinline{getWeather} (Line
4) that queries a weather forecasts service. This function is declared in Line 2. 
If the weather is clear,
Bob proposes to Alice a picnic, the cinema otherwise. 
Booking (as in Listing~\ref{es:appointment}, Lines 23-26) is needed only if 
Alice accepts the cinema proposal.

As detailed in~\cite{TR}, to obtain a deadlock-free application, we
require the code of choreographies and rules to satisfy a
well-formedness syntactic condition called
\emph{connectedness}. Intuitively, connectedness ensures that
sequences of actions are executed in the correct order
and avoids interference between parallel
interactions.
%
%
Requiring this condition does not hamper programmability, since it
naturally holds in most of the cases, and it can always be enforced
automatically via small patches to the choreography which preserve the
behaviour of the program, as discussed in~\cite{WWVlanese}.  Also,
checking connectedness is efficient, i.e., polynomial w.r.t.~the size
of the code~\cite{TR}.

%% file: implementation.tex
\section{Implementation}\label{sec:impl}
Our prototype implementation of \AIOCJ{} is composed of two
elements: the \AIOCJ{} Integrated Development Environment (IDE), 
named \toolplugin{}, and the adaptation middleware that enables 
AIOC programs to adapt, called \toolmid{}.

\toolplugin{} is a plug-in for Eclipse~\cite{eclipse} based on
Xtext~\cite{xtext}. Xtext provides features such as syntax highlighting, 
syntax checking, and code completion, which help developers in writing
choreographies and adaptation rules. Also, starting from a 
grammar, Xtext generates the parser for
programs written in the AIOC language. Result of the parsing is an
abstract syntax tree (AST) we use to implement (\emph{i}) the checker
for connectedness for choreographies and rules and
(\emph{ii}) the generation of Jolie code for each role. The
connectedness check has polynomial computational
complexity~\cite{TR} thus making it efficient enough to
be performed on-the-fly while editing the code.

The target language of code generation is Jolie~\cite{jolie}.  Jolie supports
architectural primitives such as dynamic embedding, aggregation, and
redirection that we exploit to implement the adaptation
mechanisms. Moreover, Jolie supports a wide range of communication
technologies (TCP/IP sockets, local memory, Bluetooth) and of data
formats (e.g., HTTP, SOAP, JSON). \AIOCJ{} inherits this ability.
\iftoggle{tech_report}{
Actually, any language that supports runtime code evaluation can
implement our adaptation mechanisms. However, we deem that using Jolie
architectural primitives particularly suits the requirements of our
framework.	
}{}
The compilation generates a Jolie service for each
role. The execution of scopes is delegated to sub-services accessed using
Jolie redirection facility.
Adaptation is enacted by disabling the current
sub-service and replacing it with a new one, obtained from the
adaptation server. To grant to all the sub-services access to variables, the state is 
stored by a dedicated sub-service local to the role. 
Auxiliary messages are exchanged to ensure that 
both the adaptation and the choices taken by the if and while constructs are done 
in a coordinated way.
In particular, the scope
execution not only requires interaction with the adaptation manager,
but also communications among the different roles, ensuring that they
all agree on whether adaptation is needed or not, and, in case, on
which rule to apply. Indeed, the decision is taken by the role
coordinating the adaptation and then communicated to other roles. Note
that the different roles cannot autonomously take the decision, since
if they take it at different times, changes in the environment or in the
sets of available rules may lead to inconsistent decisions.

Synchronous message exchange is implemented on top of an asynchronous 
communication middleware by a sub-service that works as a \emph{message 
handler}.
The message handler of the starter role also ensures that, before the
actual communication in the choreography starts, all the roles are
ready.

\toolmid{} is implemented in Jolie and it includes:
\begin{itemize}
  \item many, possibly distributed, \emph{adaptation servers} where
    rules are published. Adaptation servers can be deployed and
    switched on and off at runtime;
  \item an \emph{adaptation manager} that acts as a registry for
    adaptation servers and clients;
  \item an \emph{environment} service that stores and makes available
    environment information. Environment information can change at any
    moment.
\end{itemize}

When an \AIOCJ{} program reaches a scope, it queries the adaptation manager
for a rule matching that scope. The adaptation manager queries each adaptation
server sequentially, based on their order of registration. Each server checks
the applicability condition of each of its rules. The first rule whose
applicability condition holds is applied. In particular, the code of the rule
is sent to the role coordinating the adaptation (via the adaptation manager)
which distributes it to the involved roles. In each role, the new code
replaces the old one. The study of more refined policies for rule selection,
e.g., based on priorities, is a topic for future work.

%% file: validation.tex
\section{Validation}
\label{sec:validation}
In this section, we give a preliminary empirical validation of our
implementation. The main aim is to test how our mechanisms for adaptation 
impact on performances.

In the literature, to the best of our knowledge, there is no approach
to adaptation based on choreography programming. Thus, it is difficult
to directly compare our results with other existing approaches. Moreover, we 
are not aware of any established benchmark to evaluate adaptive applications. 
For this reason, we tested \AIOCJ{} performances by applying it to two 
typical programming patterns: \emph{pipes} and 
\emph{fork-joins}.
%
Since we are interested in studying the cost of adaptation, our scenarios
contain minimal computation and are particularly affected by the overhead of 
the adaptation process. Clearly, the percentage of the overhead due to adaptation will be 
far lower in real scenarios, which are usually more computationally intensive.
In the first scenario, we program a pipe executing $n$ tasks (in a
pipe, the output of task $t_{i}$ is given as input to task $t_{i+1}$,
for $i \in \{1,\ldots, n-1\}$). To keep computation to a minimum, each
task simply computes the increment function.
%
%
In the fork-join scenario, $n$ tasks are computed in parallel. Each
task processes one character of a message of length $n$, shifting it by one 
position. The message is stored in an external service.\footnote{
The code of both 
scenarios is in\iftoggle{tech_report}{
Appendix~\ref{sec:code_validation}.
}{
the companion technical report~\cite{AIOCJ_TR}.
}
}	


To enable adaptation, each task is enclosed in a scope.  We test both
scenarios with an increasing number of tasks $n \in \{10, 20,
\dots, 100\}$ to study how performances scale as the number of
adaptation scopes increases. We evaluate performances in different
contexts, thus allowing us to understand the impact of different
adaptation features, such as scopes, adaptation servers, and adaptation
rules.
%
%
%
\begin{description}
 \item[Context 1:] no scopes, no adaptation servers, no rules; 
 \item[Context 2:] each task is enclosed in a scope, no adaptation
   servers, no rules;
 \item[Context 3:] each task is enclosed in a scope, one adaptation
   server, no rules;
 \item[Context 4:] as Context 3, but now the adaptation server contains
   50 rules. Each rule is applicable to a unique scope $i$, and no
   rule is applicable to scopes with $i>50$. The rules are stored in
   random order.
 \item[Context 5:] as Context 4, but with $100$ rules, one for each
   scope.
\end{description}
%

\noindent
Each rule in Contexts 4 and 5 is applicable to one specific scope only
(through a unique property of the scope), hence when testing for 50
rules, only the first 50 scopes adapt.

\iftoggle{tech_report}{In order to reduce variability of performances, w}{W}e 
repeated every test 5 
times. We performed our tests on a machine equipped with a 2.6GHz quad-core 
Intel Core i7 processor and 16GB RAM. The machine runs Mavericks 10.9.3, Java
1.7.55, and Jolie r.2728.
\begin{figure}[t]
\centering
\includegraphics[width=\textwidth]{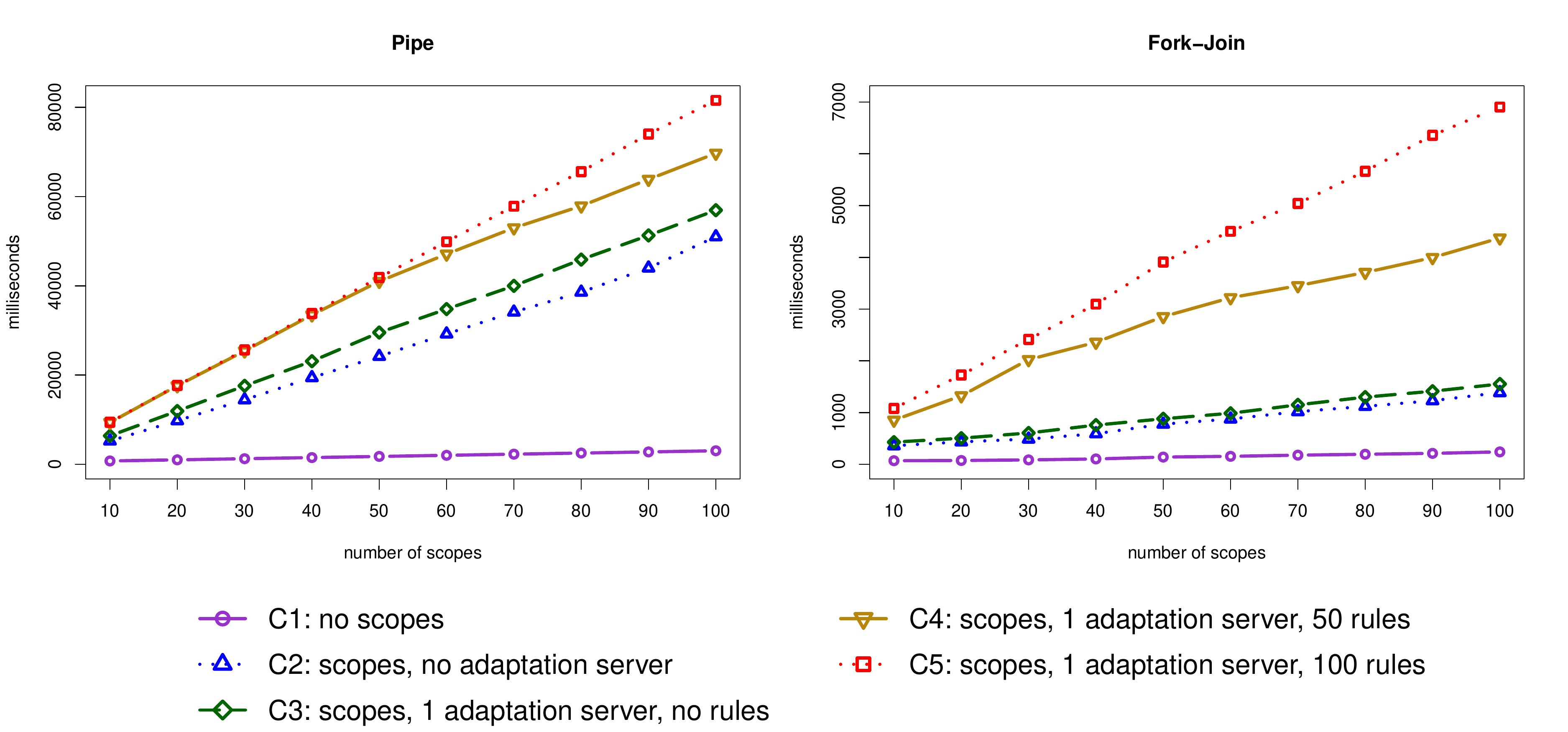}
\caption{Times of execution of the pipe (left) and the fork-join 
(right) scenarios}
\label{fig:charts}
\end{figure}
Figure~\ref{fig:charts} shows the tests for the pipe (left) 
and the fork-join (right). Both charts display on the x-axis the number of 
tasks/scopes and on the y-axis the execution time in milliseconds. 


As expected, in both scenarios there is a significant gap between
Contexts 1 and 2. In words, the introduction of scopes has a strong
effect on performances.  The ratio is 1:13 for the pipe scenario and
1:5.5 for the fork-join scenario. This is due to the auxiliary
communications needed to correctly execute a scope. 
The observed overhead is higher in the pipe scenario, since different
scopes check for adaptation in sequence, while this is done in
parallel for the fork-join scenario.
%

Adding an adaptation server (from Context 2 to Context 3) has
little impact on performances: 19\% of decay for pipe,
and 17\% for fork-join. The figures are reasonable, considered that Context 3
adds only one communication w.r.t.~Context 2.


On the contrary, there is a notable difference when adding rules to
the adaptation server (Context 4 is 1.4 times slower than Context 3 for the pipe scenario, 2.9 
for the fork-join scenario).  In Contexts 4 and 5, performances are really 
close up to 50 scopes (in the pipe scenario they almost overlap) although 
Context 5 has twice the rules of Context 4.  This illustrates that the time 
to test for applicability of rules is negligible. Hence, the highest toll on
performances is related to actual adaptation, since it requires to
transfer and embed the new code. This is particularly evident in the fork-join scenario where 
multiple adaptations are executed in parallel and the adaptation server becomes a bottleneck. This 
problem can be mitigated using multiple distributed adaptation servers.
%


The fact that the most expensive operations are scope execution and
actual adaptation is highlighted also by the results below. The table
shows the cost of different primitives, including scopes in different
contexts. Times are referred to 5 executions of the sample code 
in\iftoggle{tech_report}{
Appendix~\ref{sec:code_validation}.
}{
the companion technical report~\cite{AIOCJ_TR}.
}

%

\begin{center}
\includegraphics[width=.8\columnwidth]{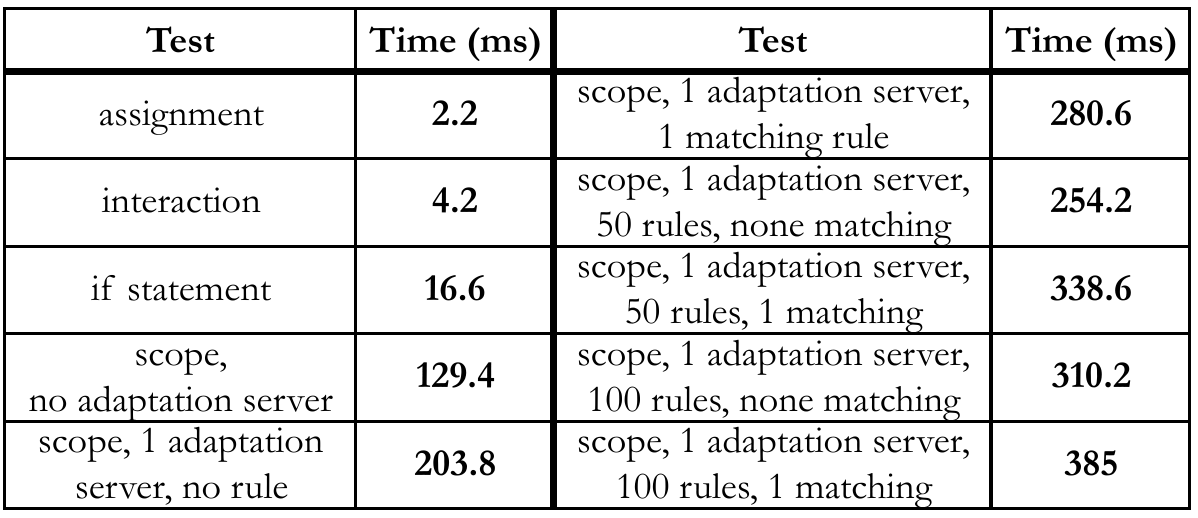}	
\end{center}
%
As future work we will exploit these results to increase the performances of
our framework, concentrating on the bottlenecks highlighted above.  For
instance, scope execution (as well as conditionals and cycles) currently
requires many auxiliary communications ensuring that all the processes agree
on the chosen path. In many cases, some of these communications are not
needed, since a process will eventually discover the chosen path
from the protocol communications. Static analysis can discover redundant 
communications and remove them. Another improvement is letting
the adaptation server send the new code directly to the involved roles, 
skipping the current forward chain.

%% file: related.tex
\section{Related Work~\& Conclusion}
\label{sec:conclusions}

This paper presented a framework for programming rule-based adaptation of
distributed applications. Its distinctive trait is that, being based on a
choreographic approach, it guarantees deadlock-freedom by construction for the
running distributed application, even in presence of adaptation rules which
were unknown when the application was started, and for any environment
condition.

Adaptation is a hot topic, and indeed there is a plethora of approaches in the
literature, see, e.g., the surveys~\cite{ADAPTGhezzi,adapt-survey12}. However,
approaches based on formal methods are only emerging recently and few of them
have been implemented in a working tool.  In particular, the use of
choreographies to capture and define adaptive applications is a novel idea.
For a discussion of works on adaptation with formal bases, but which have not
been implemented, we refer to~\cite{TR}. Here, we just recall \cite{Dezani},
which exploits a choreographic approach for self-adaptive monitoring of
distributed applications.

Among the implemented approaches, the most related to ours is
JoRBA~\cite{JORBApaper}. JoRBA features scopes and adaptation rules similar to
ours. However, JoRBA applications are not distributed and JoRBA does not
guarantee any property of the adapted application.


In~\cite{DYCHOR06} choreographies are used to propagate protocol changes to
the other peers, while~\cite{SCC09} presents a test to check whether a set of
peers obtained from a choreography can be reconfigured to match a second one.
Differently from ours, these works only provide change recommendations for
adding and removing message sequences.

Various tools
\cite{dsol-ghezzi,pistore_correct_and_adaptable,pistore_planning}
exploit automatic planning techniques in order to elaborate, at
runtime, the best sequence of activities to achieve a given goal. 
These techniques are more declarative than ours, but,
to the best of our knowledge, they 
are not guaranteed to always find a plan to adapt the application.

Among the non-adaptive languages, Chor~\cite{POPLmontesi} is the closest to
ours. Indeed, like ours, Chor is a choreographic language  that compiles to
Jolie. Actually, \AIOCJ{} shares part of the Chor code base. However, due to
the different semantics of the sequential operator and the lack of the
parallel composition in Chor, a faithful encoding of the scenarios in Section
\ref{sec:validation} is not possible, especially for the fork-join scenario.
On an almost equivalent implementation of the pipe scenario, Chor proves to be
more efficient than \AIOCJ{}.
\iftoggle{tech_report}{This performance gap reinforces our idea that  the
current implementation of \AIOCJ{} can be optimised, possibly taking inspiration
from some implementation choices made in Chor.}{}

In the future, we would like to test the expressive power of our language,
trying to encode patterns of adaptation from existing approaches. An obvious
benefit of such an encoding is that it will capture patterns of adaptation
used in real-case scenarios, guaranteeing also deadlock freedom, which is not
provided by other approaches. This task is cumbersome, due to the huge number
and heterogeneity of those approaches. Nevertheless, we already started it. In
particular, in the website~\cite{AIOCJ}, we show how to encode examples coming
from distributed~\cite{distributedAOP} and dynamic~\cite{YangCSSSM02}
Aspect-Oriented Programming (AOP) and from Context-Oriented Programming (COP)
\cite{cop}. In general, we can deal with cross-cutting concerns like logging
and authentication, typical of AOP, viewing point-cuts as empty scopes and
advices as adaptation rules. Layers, typical of COP, can instead be defined by
adaptation rules which can fire according to contextual conditions captured by
the environment. Possible extensions of our framework include the use of
asynchronous communications in the AIOC language and the introduction of
mechanisms to deal with exceptions and failures. Finally, we would like to
pursue a systematic analysis of the workflow change patterns like the  ones
presented in~\cite{change_patterns,workflow_evolution}, showing how these
patterns are captured by \AIOCJ{}.


%% file: demo.tex
\section{Demonstration} 
\label{sec:demo}
In this section, we describe, supported by screenshots, how a developer can
write an \AIOCJ{} application and its adaptation rules. We also show an
example  of execution and adaptation of the generated application.

\subsection{Writing an \AIOCJ{} application}
%
We consider the AIOC program showed in Section~\ref{sec:aioc}, but we  adapt
the deployment of external services and of the roles to run them locally.
Note that, when not explicitly defined in the \lstinline{preamble}, the
location of a role is automatically assigned to a local address.

\begin{minipage}{\textwidth}
\begin{lstlisting}[mathescape=true,numbers=left,caption=Local deployment for appointement program.,
basicstyle=\ttfamily\scriptsize ,
label=es:aioc_program,escapeinside={\%*}{*)}]
include isFreeDay from "socket://localhost:8000"
include getTicket from "socket://localhost:8001"

preamble { starter: bob }

aioc {
  // Code as in Listing %*\ref{es:appointment}*)
}
\end{lstlisting}
\end{minipage}

\begin{figure}[ht]
	\begin{center}
		\includegraphics[width=\textwidth]{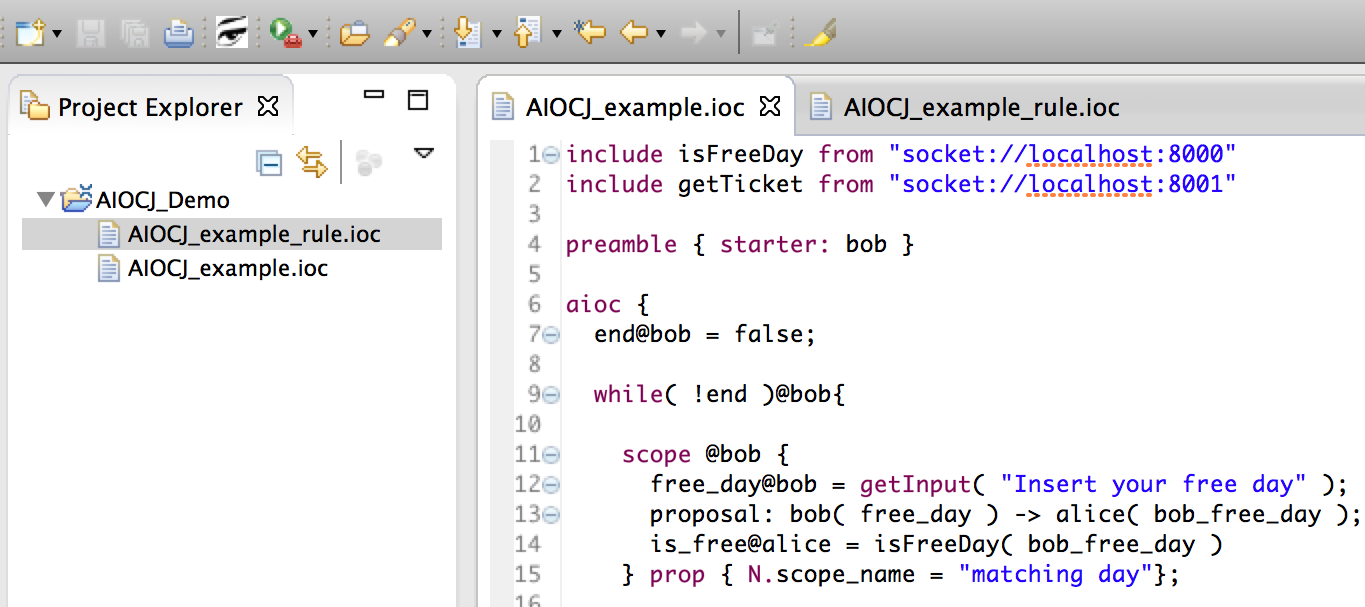}
	\end{center}
	\caption{The \AIOCJ{} eclipse plug-in}
	\label{fig:plug-in}
\end{figure}

Fig.~\ref{fig:plug-in} shows the interface of the Eclipse IDE running  the
\toolplugin{} plug-in. In \toolplugin{} developers can write choreographies
and adaptation rules. The IDE provides syntax checking and highlighting. Plus,
it performs an online check for connectedness.

\subsection{Online check for connectedness}
To see an example of how the online check for connectedness works, let us 
modify the code in Listing~\ref{es:appointment} to generate a non-connected
sequence. We switch Lines 13 and 14 of Listing~\ref{es:appointment} as
shown in the snippet below.

\begin{minipage}{\textwidth}
\begin{lstlisting}[mathescape=true,numbers=left,caption=Non-connected 
sequence.,basicstyle=\ttfamily\scriptsize,
label=es:code_con_seq,firstnumber=12]
free_day@bob = getInput( "Insert your free day" ); 
is_free@alice = isFreeDay( bob_free_day );
proposal: bob( free_day ) -> alice( bob_free_day )
\end{lstlisting}
\end{minipage}

\begin{figure}[ht]
	\begin{center}
		\includegraphics[width=0.8\textwidth]{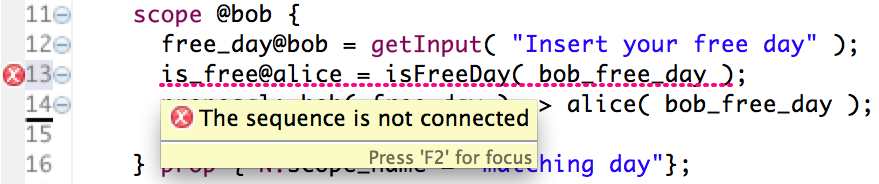}
	\end{center}
	\caption{Check for connectedness: sequence.}
	\label{fig:con_seq}
\end{figure}

Fig.~\ref{fig:con_seq} shows \toolplugin{} notifying the error on the first
non-connected instruction. Note that the code in Listing~\ref{es:code_con_seq}
does not behave as expected, since Alice uses value \lstinline{bob_free_day}
at Line 13 before receiving it at Line 14. Indeed, connectedness errors
frequently highlight errors in the logic of the application.

\begin{figure}[ht]
	\begin{center}
		\includegraphics[width=.8\textwidth]{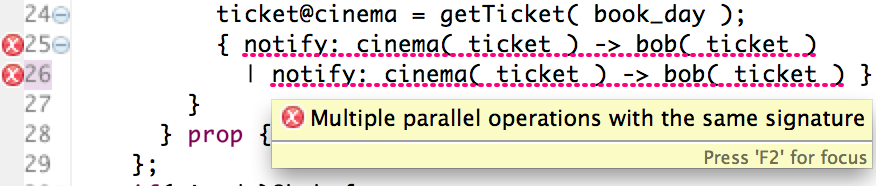}
	\end{center}
	\caption{Check for connectedness: parallel.}
	\label{fig:con_par}
\end{figure}

\noindent Similarly, \toolplugin{} also notifies when connectedness fails 
on a parallel operator (Fig.~\ref{fig:con_par}). In this case, 
w.r.t.~the code in
Listing~\ref{es:appointment}, we  miswrote the notification procedure at
Lines 25-26, sending in parallel twice the same message from cinema to
Bob. 


\subsection{Compiling and Running an \AIOCJ{} application}
\begin{figure}[ht]
	\begin{center}
		\includegraphics[width=.8\textwidth]{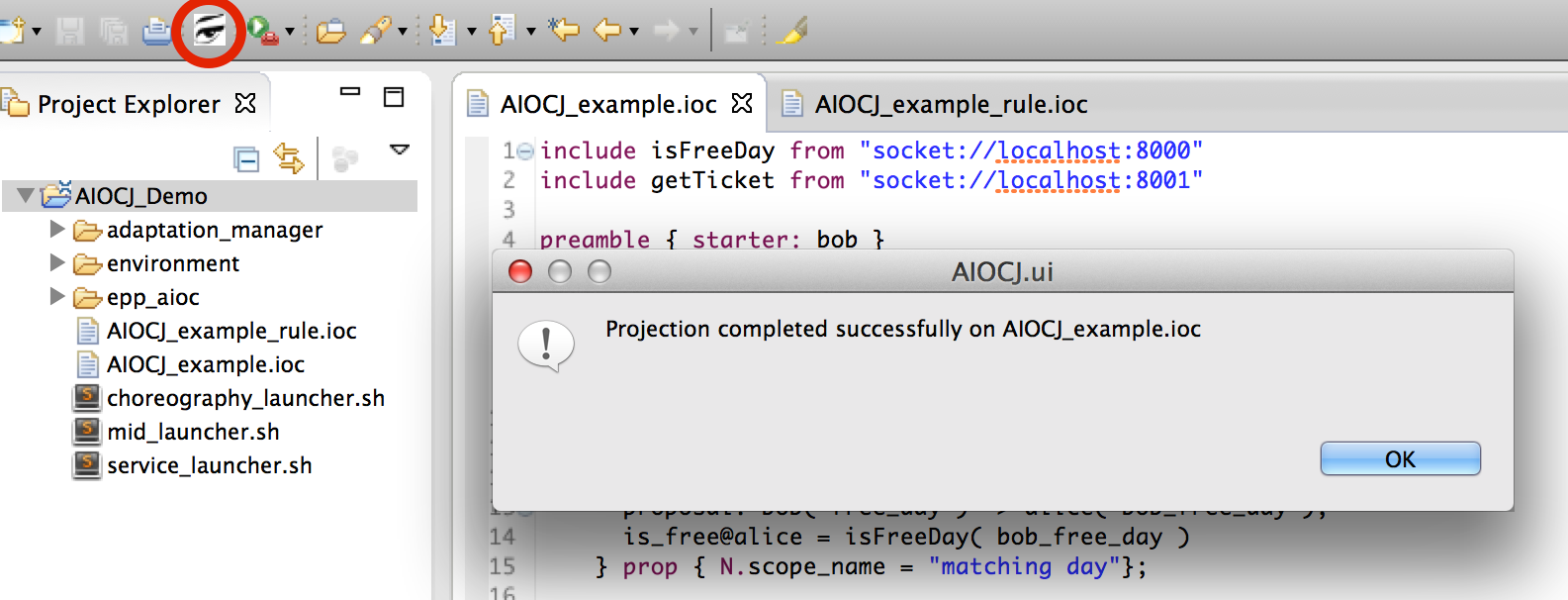}
	\end{center}
	\caption{Compilation of the \AIOCJ{} application.}
	\label{fig:aioc_comp}
\end{figure}

%
The icon
in the top toolbar of \toolplugin{} highlighted in Fig.~\ref{fig:aioc_comp}
starts the compilation of the AIOC program to Jolie. 
A dialog box notifies the successful compilation of the
program.

The compilation of an AIOC application generates three folders, as showed in
the left column of Fig.~\ref{fig:aioc_comp}. The folders contain the
adaptation manager, the environment, and the roles in the choreography (in
folder ``epp\_aioc'' ).

\toolplugin{} also projects utilities for testing the choreography locally.
The script ``choreography\_launcher.sh'' takes care of launching, in the
proper order, the processes of the adaptation middleware (the environment, the
adaptation manager, and an adaptation server, if present). Then the script
launches the roles in the choreography, beginning with the \lstinline{starter}
one.

\setJolie{}

Listing~\ref{es:bob_jolie_code} shows as an example the Jolie code generated 
from the choreography in Listing~\ref{es:appointment} for Bob.
We highlight below how some of the primitives of the AIOC language are 
encoded in Jolie.

Jolie provides \lstinline{outputPort}s for defining how services communicate. 
In our case, Bob has one \lstinline{outputPort} for each
process he communicates with (i.e., \lstinline{alice} and \lstinline{cinema}). 
Output ports are defined at Lines 1-11.

\setMyChor{}

Lines 18-21 encode the assignment \lstinline{end@bob = false}. 
At Line 18 the variable \lstinline{end} is set to 
\lstinline{false}. Then, the value is sent to the State sub-service which is 
in charge of storing it (Lines 19-21). Notice that this code is needed
since \lstinline{end} is a local variable of Bob. In contrast, the
assignment \lstinline{is_free@alice = isFreeDay( bob_free_day )} is done by
Alice and it does not appear in Bob's code.
\setJolie{}

Lines 22-30 show the encoding of the \lstinline{while} 
(Listing~\ref{es:appointment}, Line 10). First the variable \lstinline{end} 
is retrieved from the State sub-service (Line 22). 
Then the guard \lstinline{!end} is evaluated (Line 23), and the result 
of the evaluation is sent to the roles involved in the cycle (Lines 24-29). 
Finally, at Line 30, Bob enters the cycle. Notably, the sequence at Lines 
22-29 is repeated at Lines 58-65 to update the evaluation of the guard of 
the cycle. 

At Line 31, Bob enters the scope named \lstinline{"matching day"}. 
The code of the scope is not reported here since scopes are externalised as
auxiliary sub-services. However, one can see the invocation of the
sub-service at Line 35. The (blocking) invocation will return only when the 
execution of the scope is terminated.

\newpage

\begin{center}
\begin{minipage}{0.6\textwidth}
\begin{lstlisting}[mathescape=true,numbers=left,
caption=Bob Jolie code.,basicstyle=\ttfamily\scriptsize,
label=es:bob_jolie_code]
outputPort cinema {
	Location: "socket://localhost:10500/"
	Protocol: sodep
	RequestResponse: op0(OpType)(undefined)
}

outputPort alice {
	Location: "socket://localhost:10501/"
	Protocol: sodep
	RequestResponse: op0(OpType)(undefined)
}

main
{
	var12.role = "f241629d94f2";
	start@MH(var12)();
	start_bob@MH(var12)();
	end = false;
	var0.value = end;
	var0 = "end";
	set@State(var0)();
	get@State("end")(end);
	var7 = !end;
	var8.role = "709bdb0d6192";
	var8.content = var7;
	op0@cinema(var8)();
	var9.role = "709bdb0d6192";
	var9.content = var7;
	op0@alice(var9)();
	while ( var7 ){
		run@ActivityManager("0a8a09022eee")();
		var2.role = "5355f138ae39";
		get_op1@MH(var2)(var2);
		if (var2.content) {
			run@ActivityManager("c7453ecb0abf")()
		};
		get@State("end")(end);
		var6 = !end;
		if (var6) {
			showInputDialog@SwingUI("Alice refused." + 
				"Try to propose another day?[y/n]")(_r);
			var3.value = _r;
			var3 = "_r";
			set@State(var3)();
			get@State("_r")(_r);
			var5 = _r != "y";
			if (var5) {
				end = true;
				var4.value = end;
				var4 = "end";
				set@State(var4)()
			}
		};
		var10.role = "cinema";
		get_op0@MH(var10)();
		var11.role = "alice";
		get_op0@MH(var11)();
		get@State("end")(end);
		var7 = !end;
		var8.role = "709bdb0d6192";
		var8.content = var7;
		op0@cinema(var8)();
		var9.role = "709bdb0d6192";
		var9.content = var7;
		op0@alice(var9)()
	}
}
\end{lstlisting}
\end{minipage}
\end{center}
\newpage
\setMyChor{}

\begin{figure}[ht]
	\begin{center}
		\includegraphics[width=\textwidth]{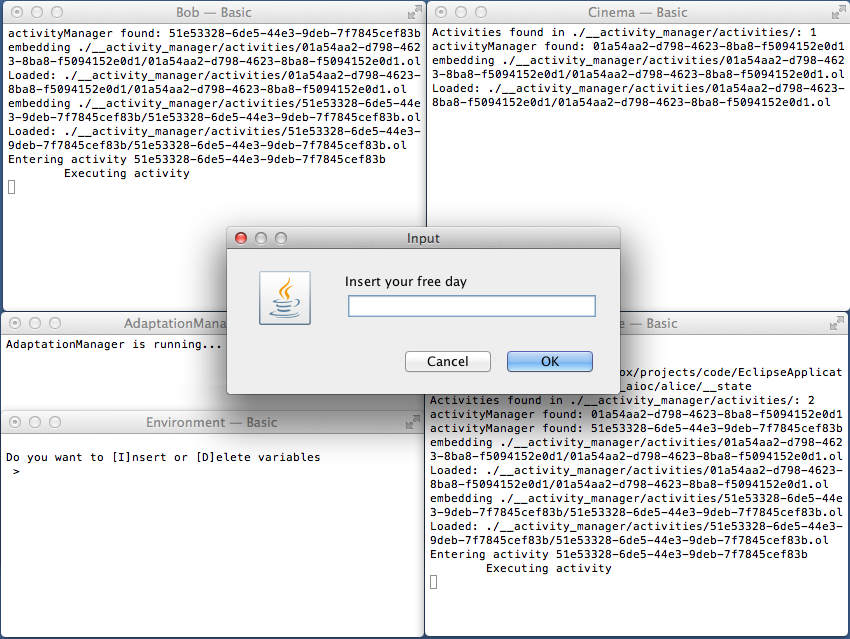}
	\end{center}
	\caption{Start of the \AIOCJ{} application.}
	\label{fig:exec_start}
\end{figure}

Fig.~\ref{fig:exec_start} shows the execution of the choreography (with one
service of the choreography per console). Consoles show also some debugging
information to clarify  the flow of execution. The application is waiting for
receiving the data that will be inserted in variable \lstinline{free_date} by
Bob (Line 12, Listing~\ref{es:appointment}).

\begin{figure}[ht]
	\begin{center}
		\includegraphics[width=\textwidth]{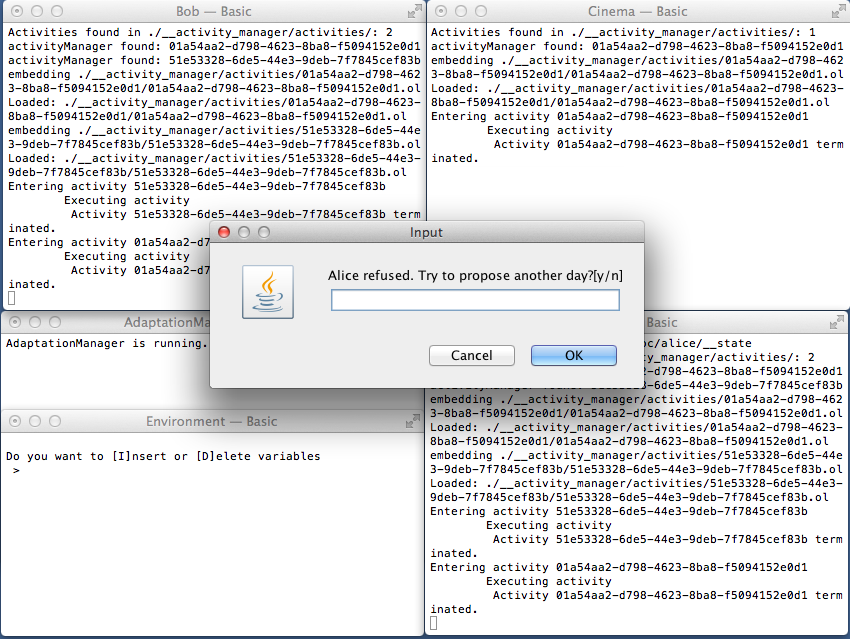}
	\end{center}
	\caption{First iteration, refusal.}
	\label{fig:exec_refusal}
\end{figure}

In this execution, Alice is free the day Bob proposes, but does not agree
with going to the cinema (Fig.~\ref{fig:exec_refusal}). Bob is notified of the
refusal and asked if he wants to retry, proposing another day.

\subsection{Writing and compiling adaptation rules}

While the \AIOCJ{} application is running, one can write, compile, and publish
new rules.

The rule in Listing~\ref{es:rule1} adapts the scope named 
\lstinline{"matching day"} at Lines 11-15 of Listing~\ref{es:appointment}. 
It changes the protocol
used by Bob to ask Alice for  the day of the appointment. Bob now first
checks, using the external function  \lstinline{hasFreeWeek}, if he is free
the whole next week. If so, he asks Alice to pick a day,  otherwise he follows
the normal protocol, suggesting to Alice a day for an appointment when he is
free.

\begin{minipage}{\textwidth}
\begin{lstlisting}[mathescape=true,numbers=left,caption=Matching day adaptation 
rule.,basicstyle=\ttfamily\scriptsize,label=es:rule1]
rule {
	include isFreeDay, hasFreeWeek from "socket://localhost:8000"
	on { N.scope_name == "matching day" }
	do {
		not_busy@bob = hasFreeWeek();
		if ( not_busy )@bob {
			free_day@alice = getInput( "Bob has a free week, 
			  insert your free day or no to refuse" );
 			if (free_day == "no")@alice {
			  is_free@alice = false;
 			} else {
			  is_free@alice = true;
			  response: alice( free_day ) -> bob( free_day ) }
 		} else {
			free_day@bob = getInput( "Insert your free day" );  
			proposal: bob( free_day ) -> alice( bob_free_day );
			is_free@alice = isFreeDay( bob_free_day ) }}}
\end{lstlisting}
\end{minipage}

Rules are compiled following the same procedure of AIOC applications. The
compilation creates a new folder ``epp\_rules'' that contains  an adaptation
server and the compiled rules.


\subsection{Adapting an \AIOCJ{} application}

\begin{figure}[ht]
	\begin{center}
		\includegraphics[width=\textwidth]{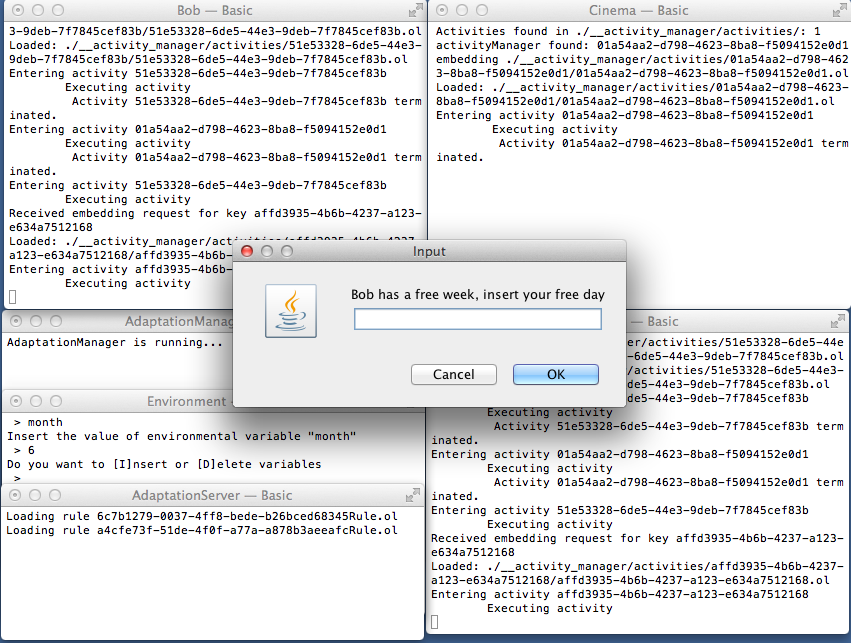}
	\end{center}
	\caption{Adaptation using the \lstinline{"matching day"} rule.}
	\label{fig:adapt_start}
\end{figure}

Inside the folder ``epp\_rules'' the script ``rules\_launcher.sh'' starts
the adaptation server. Once started, the adaptation server registers itself to
the adaptation manager. If Bob decides to proceed with a new iteration of the 
choreography, the application adapts. The rule in Listing~\ref{es:rule1} 
applies. Fig.~\ref{fig:adapt_start} shows the additional 
console where the adaptation server is running (bottom left corner). The
dialog box confirms that the rule in Listing~\ref{es:rule1} has
been applied (it notifies to Alice that Bob has a free week).

\begin{figure}[ht]
	\begin{center}
		\includegraphics[width=\textwidth]{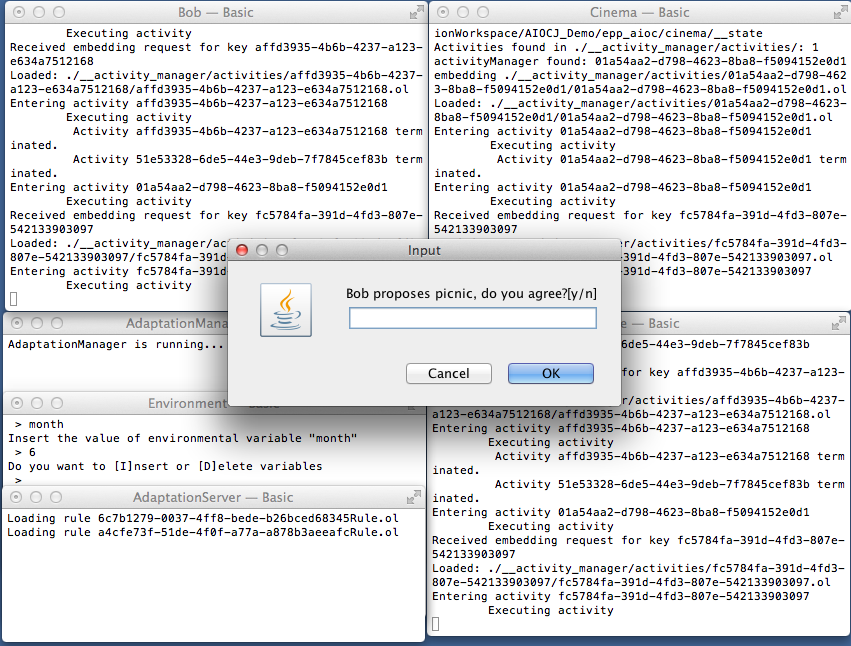}
	\end{center}
	\caption{Second iteration, adapted execution.}
	\label{fig:adapt_picnic}
\end{figure}

In order to apply also the rule in Listing~\ref{es:appointmentrule}, we set the 
variable \lstinline{month} in the environment to 6 and proceed with the choreography.
Fig.~\ref{fig:adapt_picnic} shows the result of the adaptation  in which Bob
asks whether Alice wants to go out for a picnic.

The code of the services that implement functions \lstinline{isFreeDay} and
\lstinline{getTicket} is given in~\cite{AIOCJ}, where this  example is hosted,
including the AIOC program, the rules, and the  additional services.


%% file: test_code.tex
\section{Code used for validation}
\label{sec:code_validation}

\subsection{Pipe and fork-join code} 
\label{sub:validation_code}
In this section, we provide the code of the \AIOCJ{} programs used in
Section~\ref{sec:validation}. For simplicity, we show the programs fixing the number of tasks $n = 
5$, and showing just two adaptation rules per scenario.

\noindent\textbf{Pipe scenario.}
In the pipe scenario each task computes the increment function, passing the
output of its computation as input to the next task.

\begin{minipage}{\textwidth}
\begin{lstlisting}[mathescape=true,numbers=left,
caption=Pipe scenario without scopes.,
basicstyle=\ttfamily\scriptsize ,
label=es:pipe_code_tasks]
include startTimer, endTimer from "socket://localhost:8000"
preamble{ starter: a }

aioc {
	{ x@a = 0 | x@b = 0 };
	_r@a = startTimer( n );
  x@a = 1 + x; pass: a( x ) -> b( x ); 
  x@b = 1 + x; pass: b( x ) -> a( x ); 
  x@a = 1 + x; pass: a( x ) -> b( x ); 
  x@b = 1 + x; pass: b( x ) -> a( x ); 
  x@a = 1 + x; pass: a( x ) -> b( x );
  _r@a = stopTimer( n )
}
\end{lstlisting}
\end{minipage}

Listing~\ref{es:pipe_code_tasks} shows the 5 tasks that increment variable
\lstinline{x}. After each increment the variable is sent from role \lstinline{a} to role 
\lstinline{b} or vice versa. Lines 6 and
12 are calls to the Timer service (Line 1), used to log the times of
execution.

\begin{minipage}{\textwidth}
\begin{lstlisting}[mathescape=true,numbers=left,
caption=Pipe scenario with scopes.,
basicstyle=\ttfamily\scriptsize ,
label=es:pipe_code_scopes]
include startTimer, endTimer from "socket://localhost:8000"
preamble{ starter: a }

aioc {
	{ x@a = 0 | x@b = 0 };
	_r@a = startTimer( n );
  scope @a {x@a = 1 + x; pass: a( x ) -> b( x )} prop {N.x = 1};
  scope @b {x@b = 1 + x; pass: b( x ) -> a( x )} prop {N.x = 2};
  scope @a {x@a = 1 + x; pass: a( x ) -> b( x )} prop {N.x = 3};
  scope @b {x@b = 1 + x; pass: b( x ) -> a( x )} prop {N.x = 4};
  scope @a {x@a = 1 + x; pass: a( x ) -> b( x )} prop {N.x = 5};
  _r@a = stopTimer( n )
}
\end{lstlisting}
\end{minipage}

Listing~\ref{es:pipe_code_scopes} shows the 5 scopes enclosing the tasks
of Listing~\ref{es:pipe_code_tasks}. Notably, each scope has a unique
property \lstinline{N.x} assigned incrementally. This property is used to
identify uniquely each scope.

\noindent\textbf{Rules of the pipe scenario.}
Listing~\ref{es:pipe_code_rules} shows two rules used in the pipe scenario.
The adapted behaviour increments the variable \lstinline{x} by 2.
In particular, the rule at Lines 1-4 applies to the scope at Line 7 in 
Listing~\ref{es:pipe_code_scopes}, while the subsequent rule (Lines
5-8) applies to the scope at Line 8.

\begin{minipage}{\textwidth}
\begin{lstlisting}[mathescape=true,numbers=left,
caption=Rules for pipe scenario.,
basicstyle=\ttfamily\scriptsize ,
label=es:pipe_code_rules]
rule {
  on { N.x == 1 }
  do { x@a = 2 + x; pass: a( x ) -> b( x ) }
}
rule {
  on { N.x == 2 }
  do { x@b = 2 + x; pass: b( x ) -> a( x ) }
}
\end{lstlisting}
\end{minipage}

\noindent\textbf{Fork-join scenario.} In the fork-join scenario tasks run in
parallel. Each task takes one character of a message and replaces it with
the next character in the alphabet. The program makes use of an external
service to retrieve the character of the message, get the next letter, 
and store the new character in the message.

\begin{minipage}{\textwidth}
\begin{lstlisting}[mathescape=true,numbers=left,
caption=Fork-join scenario without scopes.,
basicstyle=\ttfamily\scriptsize ,
label=es:fork_code_tasks]
include start, end from "socket://localhost:8000"
include getNthChar, getNext, setNthChar from "socket://localhost:8001"

preamble{	starter: a }

aioc {
  _r@a = start( n );
  {
     {  l0@a = getNthChar( 0 ); l0@a = getNext( l0 ); _r@a = setNthChar( 0, l0 ) }
  |  {  l1@b = getNthChar( 1 ); l1@b = getNext( l1 ); _r@b = setNthChar( 1, l1 ) }
  |  {  l2@a = getNthChar( 2 ); l2@a = getNext( l2 ); _r@a = setNthChar( 2, l2 ) }
  |  {  l3@b = getNthChar( 3 ); l3@b = getNext( l3 ); _r@b = setNthChar( 3, l3 ) }
  |  {  l4@a = getNthChar( 4 ); l4@a = getNext( l4 ); _r@a = setNthChar( 4, l4 ) }
  };
  _r@a = end( n )
}
\end{lstlisting}
\end{minipage}

Listing~\ref{es:fork_code_tasks} shows the 5 tasks of the fork-join scenario.
Each task uses a dedicated variable (\lstinline{l0, l1, ..., ln}) to ensure
thread-safety of the computation.

\begin{minipage}{\textwidth}
\begin{lstlisting}[mathescape=true,numbers=left,
caption=Fork-join scenario with scopes.,
basicstyle=\ttfamily\scriptsize ,
label=es:fork_code_scopes]
include start, end from "socket://localhost:8000"
include getNthChar, getNext, setNthChar from "socket://localhost:8001"

preamble{	starter: a }

aioc {
	_r@a = startTimer( n ); {
    scope @a {  
      l0@a = getNthChar( 0 ); l0@a = getNext( l0 ); _r@a = setNthChar( 0, l0 ) } 
      prop { N.char = 0 } |
    scope @b {  
      l1@b = getNthChar( 1 ); l1@b = getNext( l1 ); _r@b = setNthChar( 1, l1 ) } 
      prop { N.char = 1 } |
    scope @a {  
      l2@a = getNthChar( 2 ); l2@a = getNext( l2 ); _r@a = setNthChar( 2, l2 ) } 
      prop { N.char = 2 } |
    scope @b {  
      l3@b = getNthChar( 3 ); l3@b = getNext( l3 ); _r@b = setNthChar( 3, l3 ) } 
      prop { N.char = 3 } |
    scope @a {  
      l4@a = getNthChar( 4 ); l4@a = getNext( l4 ); _r@a = setNthChar( 4, l4 ) } 
      prop { N.char = 4 } 
  };
  _r@a = stopTimer( n )
}
\end{lstlisting}
\end{minipage}

Likewise the pipe scenario, the scopes in Listing~\ref{es:fork_code_scopes}
enclose the tasks of Listing~\ref{es:fork_code_tasks}. Also in this scenario
each scope has a unique property \lstinline{N.char} that identifies it 
uniquely.

\noindent\textbf{Rules of the fork-join scenario.} The adapted behaviour introduces the
new function \lstinline{GetDoubleNext} that gets a character and returns the
character two positions after it in the alphabet. Listing~\ref{es:fork_code_rules}
shows two rules used in the fork scenario. The rule at Lines 1-6 applies to
the scope at Lines 8-10 of Listing~\ref{es:fork_code_scopes}, while the rule
at Lines 7-12 applies to the scope at Lines 11-13.

\begin{minipage}{\textwidth}
\begin{lstlisting}[mathescape=true,numbers=left,
caption=Rules of pipe scenario.,
basicstyle=\ttfamily\scriptsize ,
label=es:fork_code_rules]
rule {
include getNthChar, getDoubleNext, setNthChar from "socket://localhost:8001"
  on { N.char == 0 } 
  do {  l0@b = getNthChar( 0 ); l0@b = getDoubleNext( l0 ); 
    _r@b = setNthChar( 0, l0 ) }
}
rule {
include getNthChar, getDoubleNext, setNthChar from "socket://localhost:8001"
  on { N.char == 1 }
  do {  l1@b = getNthChar( 1 ); l1@b = getDoubleNext( l1 ); 
    _r@b = setNthChar( 1, l1 ) }
}
\end{lstlisting}
\end{minipage}


\subsection{\AIOCJ{} programs used for benchmarking primitives} 
\label{sub:primitives}
\begin{lstlisting}[mathescape=true,numbers=left,
caption={Code for benchmarking assignment, interaction, and the if statement}.,
basicstyle=\ttfamily\scriptsize,
label=es:innertimes1]
include start, end from "socket://localhost:8000"
preamble { starter: a }
aioc {
  _r@a = start( "assignment" );
  x@a = 1;
  _r@a = end( "assignment" );
  
  _r@a = start( "interaction" );
  pass: a( x ) -> b( x );
  _r@a = end( "interaction" );
  
  _r@a = start( "if statement" );
  if ( x == 1 )@a { skip };
  _r@a = end( "if statement" )
}
\end{lstlisting}
\hspace{1em}
\begin{lstlisting}[mathescape=true,numbers=left,
caption={Code for benchmarking the scope primitive}.,
basicstyle=\ttfamily\scriptsize,
label=es:innertimes2]
include start, end 
  from "socket://localhost:8000"

preamble { starter: a }

aioc {
  _r@a = start( "scope" );
  scope @a{ skip } 
  prop { N.scope_name = "applicable"};
  _r@a = end( "scope" )
}
\end{lstlisting}

Listings~\ref{es:innertimes1} and~\ref{es:innertimes2} show the code used 
to measure the performances of primitives of \AIOCJ{}. 
We tested the performances of the scope (Listing~\ref{es:innertimes2}) under 
7 contexts:

\begin{enumerate}
  \item without adaptation servers;
  \item with 1 adaptation server, but without rules;
  \item with 1 adaptation server and 1 matching rule;
  \item with 1 adaptation server, 50 rules but none matching;
  \item with 1 adaptation server, 50 rules, and 1 matching;
  \item with 1 adaptation server, 100 rules but none matching;
  \item with 1 adaptation server, 100 rules, and 1 matching;
\end{enumerate}

\begin{minipage}{\textwidth}
\begin{lstlisting}[mathescape=true,numbers=left,
caption={Rules used for benchmarking scopes}.,
basicstyle=\ttfamily\scriptsize,
label=es:innertimes2_rules]
rule {
  on { N.scope_name == "applicable" }
  do { skip }
}
rule {
  on { N.scope_name == "non-applicable" }
  do { skip }
}
\end{lstlisting}
\end{minipage}

Listing~\ref{es:innertimes2_rules} shows the two kinds of rules used for 
testing performances of Listing~\ref{es:innertimes2}. The first one (Lines 1-4) 
is the matching rule used in contexts 3, 5, and 7. The second one (Lines 5-8) is 
a non matching rule present, in different copies, in contexts 4, 5, 6, and 7.
